\documentclass[preprint,12pt]{elsarticle}

\usepackage{graphicx}

\usepackage{amssymb, amsthm}

\biboptions{compress}

\journal{\hspace{-3.5cm} \raisebox{-1mm}{\begin{tikzpicture} \draw [fill = white, white] (0, 0) rectangle (3.5, 0.4); \end{tikzpicture}}}

\usepackage{amsmath}
\usepackage{url}
\usepackage{color}
\usepackage{tensor}
\usepackage{mathrsfs}
\usepackage{nicefrac}
\usepackage{tikz-cd}
\usepackage{tikz}
\usetikzlibrary{decorations.markings}
\usetikzlibrary{matrix,arrows}
\usepackage{array}
\usepackage{ulem}
\usepackage{placeins}

\usepackage{pdflscape}

\usepackage[latin1]{inputenc}      
\usepackage[T1]{fontenc}    
\usepackage{ae,aecompl}                          
\usepackage{dsfont}
\usepackage{appendix}
\usepackage{amsfonts,amsmath,latexsym,bbm}
\usepackage{amssymb}
\usepackage{accents}
\usepackage{graphics}
\usepackage{graphicx}
\usepackage{booktabs,longtable}
\usepackage{bm}
\usepackage{dcolumn}
\usepackage{enumitem}
\usepackage{upgreek}
\usepackage{tensor}
\usepackage{caption}

\usepackage{tikz}
\usepackage{tikz-cd}
\usepackage{tikz-3dplot}
\usetikzlibrary{decorations.markings}
\usetikzlibrary{matrix,arrows}
\usetikzlibrary{calc}
\usetikzlibrary{shapes}
\usetikzlibrary{positioning}
\usetikzlibrary{fit}

\usepackage{pgfplots}
\pgfplotsset{width=7cm,compat=1.5.1}

\usetikzlibrary{decorations.pathreplacing}
\makeatletter
\def\underbrace#1{%
  \@ifnextchar_{\tikz@@underbrace{#1}}{\tikz@@underbrace{#1}_{}}}
\def\tikz@@underbrace#1_#2{%
  \tikz[baseline=(a.base)] {\node[inner sep=2] (a) {\(#1\)};
  \draw[thick,line cap=round,decorate,decoration={brace,amplitude=4pt}]
    (a.south east) -- node[pos=0.5,below,inner sep=7pt] {\(\scriptstyle #2\)} (a.south west);}}
\makeatother

\usepackage[linktocpage=true]{hyperref}

\makeatletter
\renewcommand*{\eqref}[1]{%
  \hyperref[{#1}]{\textup{\tagform@{\ref*{#1}}}}%
}
\makeatother

\setlist{font=\normalfont\itshape} 

\renewcommand{\vec}[1]{\boldsymbol{#1}}
\newcommand{\tsr}[1]{\overset\leftrightarrow{#1}}
\newcommand{\ext}{_{\rm ext}}
\newcommand{\tot}{_{\rm tot}}
\newcommand{\ind}{_{\rm ind}}

\newcommand{\h}{\hspace{1pt}}
\newcommand{\hh}{\hspace{0.5pt}}
\newcommand{\mh}{\hspace{-1pt}}

\newcommand{\de}{\mathrm d}

\newcommand{\lar}[1]{\textnormal{\mbox{\large $#1$}}}
\renewcommand{\i}{\mathrm i}

\definecolor{oldgray}{gray}{0.4}

\newcommand{\T}{_{\mathrm T}}
\renewcommand{\L}{_{\mathrm L}}
\newcommand{\LL}{_{\mathrm{LL}}}
\newcommand{\LT}{_{\mathrm{LT}}}
\newcommand{\TL}{_{\mathrm{TL}}}
\newcommand{\TT}{_{\mathrm{TT}}}

\DeclareMathAlphabet{\mathbbmsl}{U}{bbm}{m}{sl}

\numberwithin{equation}{section}

\begin{document}

\begin{frontmatter}



\title{General form of the full electromagnetic Green function in materials physics}


\author[heidelberg,aachen]{G.A.H.~Schober\corref{cor1}}
\ead{schober@physik.rwth-aachen.de}

\author[freiberg]{R.~Starke}
\ead{Ronald.Starke@physik.tu-freiberg.de}

\cortext[cor1]{Corresponding author.}

\address[heidelberg]{Institute for Theoretical Physics, Heidelberg University, \\ Philosophenweg 19, 69120 Heidelberg, Germany \vspace{0.1cm}}
\address[aachen]{Institute for Theoretical Solid State Physics, RWTH Aachen University, Otto-Blumenthal-Stra\ss e 26, 52074 Aachen \vspace{0.1cm}}
\address[freiberg]{Institute for Theoretical Physics, TU Bergakademie Freiberg, \\ Leipziger Stra\ss e 23, 09596 Freiberg, Germany}

\begin{abstract}
In this article, we present the general form of the full electromagnetic Green function which is suitable for the application in bulk materials physics. In particular, we show
how the seven adjustable parameter functions of the free Green function translate into seven corresponding parameter functions of the full Green function.
Furthermore, for both the fundamental response tensor and the electromagnetic Green function,
we discuss the reduction of the Dyson equation on the four-dimensional Minkowski space
to an equivalent, three-dimensional Cartesian Dyson equation.
\end{abstract}


\end{frontmatter}



\newpage
\tableofcontents
\newpage

\section{Introduction}\label{sec:intro}

The full electromagnetic Green function with the ensuing relation to its free counterpart via the famous {\it Dyson equation} is a well-established object of study in {\it quantum electrodynamics} 
(see the standard textbooks \cite[Eq.~(10.5.14)]{Weinberg1}, \cite[\S\,7.1.1]{Itzykson}, \cite[\S\,7.5]{Peskin}, \cite[p.~175, Problem 9.2 or Eq.~(13.32)]{Huang}, or \cite[Eqs.~(9.87)--(9.88)]{Ryder}).
By contrast, in {\it ab initio materials physics} one typically restricts attention to the so-called screened potential \cite{Hedin65,Hedin69}, 
such that the relativistic {\it Schwinger-Dyson equations} \cite[\S\,10.1]{Itzykson} 
can be reduced to their non-relativistic counterpart, the so-called {\it Hedin equations} \cite{Aryasetiawan98,Schindl}. However, 
it first became obvious in {\it plasma physics} that the full electromagnetic Green function is 
a natural object to study in {\it electrodynamics of materials} as well (plasmas in this case), 
especially when it comes to the formulation of wave equations in media \cite[\S\,2.1]{Melrose1Book}.

This line of research has been resumed by the authors of the present article in their quest for a microscopic formulation of electrodynamics in media,
which is both Lorentz covariant \cite{EDOhm, EDLor} and in accordance with the common practice in ab initio materials physics \cite{ED2}.
In particular, it turned out that in condensed matter physics, the wave equation can be reformulated concisely in terms
of the full electromagnetic Green function \cite[\S\,4.1.4]{Refr}. Not surprisingly in this context, a 
succinct connection between the {\it Cartesian} dielectric tensor and the {\it spatial} part of the full electromagnetic Green function in the temporal
gauge has been unearthed \cite[Eq.~(3.44)]{EDWave}. As this relation crucially hinges on the gauge condition for
the Green function, these findings make it desirable to study the most general form of the full electromagnetic Green function
in bulk materials, a problem which had already been solved for the {\it free} electromagnetic Green function in Ref.~\cite[\S\,3.3]{ED1}. 

Here, we address this issue as follows: After introducing some technicalities in \S\,\ref{Sec:Review}
together with a short review of the free Green function in \S\,\ref{sec_ClassED}, the basics of microscopic electrodynamics
in materials---including, in particular, the fundamental response tensor---are introduced in \S\,\ref{sec_fund_resp}.
The central \S\,\ref{sec_femGF} then introduces the full Green function, derives its most general form, and
discusses the reduction of the corresponding four-dimensional Dyson equation to its three-dimensional Cartesian version.

\section{Basic techniques} \label{Sec:Review}

\subsection{Projector formalism} \label{sec_proj}

In this first subsection, we shortly assemble some technicalities which~will be useful in the following.
The {\itshape Minkowskian longitudinal and transverse projectors} are operators acting on the four-dimensional Minkowski space as follows (see \cite[\S\,3.3]{ED1}):
\begin{align}
 (P_{\rm L})\indices{^\mu_\nu}(k) & = \frac{k^\mu k_\nu}{k^2} \,, \label{Mink_long-proj} \\[5pt]
 (P_{\rm T})\indices{^\mu_\nu}(k) & = \eta\indices{^\mu_\nu} - \frac{k^\mu k_\nu}{k^2} \,,
\end{align}
where $k^{\mu} = (\omega/c, \h \vec k)^{\rm T}$ denotes a four-wavevector,
%
%
$\eta_{\mu\nu} = \mathrm{diag}(-1,\,1,\,1,\,1)$ the Minkowski metric, and $k^2 = k^\mu k_\mu = -\omega^2/c^2 + |\vec k|^2$\hh.
These operators being given, any Minkowski four-vector field $f(k) = f^\mu(k)$ can be uniquely decomposed into its longitudinal and transverse contributions,
\begin{equation}
f(k) = f\L(k) + f\T(k) \,,
\end{equation}
where $f\L(k) = P\L(k) \h f(k)$ and $f\T(k) = P\T(k) \h f(k)$. Similarly, any
 $(4 \times 4)$ {\itshape Minkowski tensor} $M(k) = M\indices{^\mu_\nu}(k)$ can be uniquely decomposed into four contributions,
\begin{equation}
 M(k) = M_{\rm LL}(k) + M_{\rm LT}(k) + M_{\rm TL}(k) + M_{\rm TT}(k) \,,
\end{equation}
which are respectively defined as
\begin{align}
 M_{\rm LL}(k) = P\L(k) \h M(k) \h P\L(k) \,, \\[3pt]
 M_{\rm LT}(k) = P\L(k) \h M(k) \h P\T(k) \,, \\[3pt]
 M_{\rm TL}(k) = P\T(k) \h M(k) \h P\L(k) \,, \\[3pt]
 M_{\rm TT}(k) = P\T(k) \h M(k) \h P\T(k) \,.
\end{align}
In particular, a four-vector $f$ is called Minkowski-transverse if
\begin{equation}
f\L = 0 \quad \textnormal{and} \quad f\T = f \,,
\end{equation}
and a Minkowski tensor $M$ is called Minkowski-transverse if
\begin{equation}
M_{\rm LL} = M_{\rm LT} = M_{\rm TL} = 0 \quad \textnormal{and} \quad M_{\rm TT} = M \,.
\end{equation}
By contrast, the {\itshape Cartesian longitudinal and transverse projectors} are defined 
as operators acting on the three-dimensional space \cite[\S\,2.1]{ED1}, i.e.,
\begin{align}
 (P_{\mathrm L} )_{ij}(\vec k) & = \frac{k_i \hh k_j}{|\vec k|^2} \,, \label{eq_defPL} \\[3pt]
 (P_{\mathrm T} )_{ij}(\vec k) & = \delta_{ij} - \frac{k_i \hh k_j}{|\vec k|^2} \,. \label{eq_defPT}
\end{align}
For details on these operators, see Refs.~\cite[\S\,2.1]{EDWave} or \cite[\S\,2.1]{ED1}.

\subsection{(3+1)-formalism} \label{3plus1}

Although the theory of relativity is in principle completely symmetric with respect to space and time,
it is useful for many purposes to formally break this manifest symmetry by decomposing any Minkowski four-vector
into its temporal and spatial components. This is accomplished by the so-called $(3+1)$-formalism 
(see Refs.~\cite[\S\,3.3]{Misner} and \cite{Alcubierre,GambiniPullin} for applications in general relativity, Refs.~\cite{EDOhm, EDLor} for applications in condensed matter physics).
In this subsection, we shortly explain some aspects of this formalism as far as they are needed for the purposes of this article.
First, we introduce the dimensionless Cartesian~vector \smallskip
\begin{equation} \smallskip
 \vec u := \frac{c \hh \vec k}{\omega} \,, \smallskip
\end{equation}
and the analogous Minkowski four-vector
\begin{equation}
 u^\mu := \frac{c \hh k^\mu}{\omega} \,.
\end{equation}
Then, we can write the contravariant four-vector $u^\mu$ as a $(4 \times 1)$-matrix,
\begin{equation} \label{zwischen_5}
 u^\mu = \left( \!\! \begin{array}{cc} 1 \\[2pt] \vec u \end{array} \!\!  \right) \,,
\end{equation}
and the corresponding covariant four-vector as a $(1 \times 4)$-matrix,
\begin{equation}
 u_\nu = \big({-1}, \h \vec u^{\rm T}\hh\big)\,.
\end{equation}
Furthermore, the projector \eqref{Mink_long-proj} can be written in the $(3+1)$-formalism as
\begin{equation}
 (P\L)\indices{^\mu_\nu} = \frac{1}{|\vec u|^2 -1} \begin{pmatrix} -1 & \vec u^{\rm T} \\[5pt] -\vec u & \vec u \vec u^{\rm T} \end{pmatrix}, \smallskip
\end{equation}
or equivalently as
\begin{equation}
 (P\L)\indices{^\mu_\nu} =  \frac{1}{|\vec u|^2 - 1}  \left( \!\! \begin{array}{cc} 1 \\[2pt] \vec u \end{array} \!\!  \right) \mh \big({-1}, \h \vec u^{\rm T} \hh \big) \,.
\end{equation}
For later purposes, we also introduce the $(4 \times 3)$-matrix
\begin{equation} \label{def_mat_1}
 \bigg( \!\mh \begin{array}{c} \vec u^{\rm T} \\[2pt] \tsr 1 \end{array} \!\mh \bigg) = \left( \!\mh \begin{array}{ccc} u_1 & u_2 & u_3 \\[2pt] 1 & 0 & 0 \\[2pt] 0 & 1 & 0 \\[2pt] 0 & 0 & 1 \end{array} \!\mh \right),
\end{equation}
and the $(3\times 4)$-matrix
\begin{equation} \label{def_mat_2}
 \big({-\vec u}, \h \tsr 1 \h \big) = \left( \! \begin{array}{cccc} -u_1 & 1 & 0 & 0 \\[2pt] -u_2 & 0 & 1 & 0 \\[2pt] -u_3 & 0 & 0 & 1 \end{array} \! \right).
\end{equation}
These auxiliary objects will become relevant in \S\,\ref{subsec_penultimate}.

\section{Free electromagnetic Green function} \label{sec_ClassED}

\subsection{Definition}\label{subsec_freeGF}

As a matter of principle, the fundamental equation of motion for the four-potential $A^\nu(x)$ generated by the four-current $j^\mu(x)$ reads
\begin{equation}
\left(\eta\indices{^\mu_\nu}\Box+\partial^\mu\partial_\nu\right) \mh A^\nu(x)=\mu_0 \h j^\mu(x) \,, \label{fund_wave_eq}
\end{equation}
where \h$\Box = -\partial_\lambda \h \partial^\lambda$ is called the d'Alembert operator.
A particular solution of this equation is given in terms of the (tensorial) free electromagnetic Green function $D_0$ by
\begin{equation}
 A^\nu(x) = \int \! \de^4 x' \, (D_0)\indices{^\nu_\lambda}(x - x') \, j^\lambda(x') \,,
\end{equation}
or in Fourier space by
\begin{equation}
A^\nu(k)=(D_0)\indices{^\nu_\lambda}(k) \, j^\lambda(k)\,. \label{eq_introD_0}
\end{equation}
Here, the free Green function fulfills {\itshape per definitionem} \cite[\S\,3.3]{ED1} the equation
\begin{equation} \label{def_freeGF}
 \left(\eta\indices{^\mu_\nu}\Box+\partial^\mu\partial_\nu\right) (D_0)\indices{^\nu_\lambda}\, j^\lambda = \mu_0  \h j^\mu 
\end{equation}
for any {\it physical} four-current, i.e., for any four-current which satisfies the continuity equation, \smallskip
\begin{equation}
 \partial_\mu \h j^\mu = 0 \,, \smallskip \label{eq_continuity}
\end{equation}
and which is hence Minkowski-transverse,
\begin{equation}
P\T \h j = j \,.
\end{equation}
On the other hand, the equation of motion \eqref{fund_wave_eq} can be rewritten as
\begin{equation}
 \Box \h P\T \h A = \mu_0 \h j = \mu_0 \h P\T \h j \,,
\end{equation}
and correspondingly, we obtain the defining equation for the (tensorial) electromagnetic Green function in the form \smallskip
\begin{equation}
 \Box \h P\T \h D_0 \h P\T = \mu_0 \h P\T \,. \label{eq_defEqFGF} \smallskip
\end{equation}
We note that this defining equation for the Green function is less restrictive
than the one used by D.\,B.~Melrose \cite[Eq.~(2.1.7)]{Melrose1Book}; together with his Eq.~(2.1.5), however, this approach does not allow one to identify the essential free parameters of the free Green function.
In the following subsection, we will discuss the general solution of the above Eq.~\eqref{eq_defEqFGF}, which at the same time provides the general form of the free electromagnetic Green function.

\subsection{General form} \label{free_GF_mink}
\subsubsection{Projector formalism}\label{subsec_genformGF}

As has been shown in Ref.~\cite[\S\,3.3]{ED1}, the most general form of the free electromagnetic Green function reads
\begin{equation} \label{eq_gen_GF_dimless}
 (D_0)\indices{^\mu_\nu}(k) = \mathbbmsl D_0(k) \mh \left( \eta\indices{^\mu_\nu} + \frac{c k^\mu}{\omega} \h f_\nu(k) + g^\mu(k) \h \frac{c k_\nu}{\omega} + \frac{c k^\mu}{\omega} \h h(k) \h \frac{c k_\nu}{\omega} \right),
\end{equation}
where $\mathbbmsl D_0(k)$ denotes the (scalar) Green function of the d'Alembert operator in Fourier space \cite[\S\,3.1]{ED1},
\begin{equation} \label{defscalarD}
\mathbbmsl D_0(\vec k, \omega) = \frac{\mu_0}{-\omega^2/c^2+ |\vec k|^2} \,,
\end{equation}
and the complex parameter functions $f_\nu$, $g^\mu$ and $h$ can be chosen arbitrarily up to the constraints of Minkowski-transversality, i.e.,
\begin{equation} \label{mink_trans}
 f_\nu(k) \h\hh k^\nu = k_\mu \h g^\mu(k) = 0 \,.
\end{equation}
In special cases, further restrictions on these parameter functions may be derived. For example,  in the context of condensed matter theory
one thinks of the Green function as a {\it retarded} response function, which satisfies the Kramers--Kronig relations between real and imaginary parts. Furthermore, the retarded Green function is definitely real-valued implying
\begin{equation}
 (D_0)^{\mu\nu}(k) = (D^*_0)^{\mu\nu}(-k) \,,
\end{equation}
from which we obtain the relations
\begin{align*}
 f^\mu(k) & = f^{*\mu}(-k) \,, \\[5pt]
 g^\mu(k) & = g^{*\mu}(-k) \,, \\[5pt]
 h(k) & = h^*(-k) \,.
\end{align*}
On the other hand, in quantum electrodynamics the electromagnetic Green function is identified with the {\it time-ordered}
expectation value (using conventions as in Ref.~\cite[Eq.~(E.17)]{EffWW})
\begin{equation}
{-\i} \hh \hbar \h c \, (D_0)^{\mu\nu}(x, x')=\langle \h\mathcal T\hat A^\mu(x) \h \hat A^\nu(x') \rangle\,.
\end{equation}
Since the field operators in the range of the time-ordering operator commute with each other, this implies the additional relation
\begin{equation}
 (D_0)^{\mu\nu}(x, x') = (D_0)^{\nu\mu}(x', x) \,,
\end{equation}
which in Fourier space translates into
\begin{equation}
 (D_0)^{\mu\nu}(k) = (D_0)^{\nu\mu}(-k) \,.
\end{equation}
Imposing this condition on the Green function in its general form, Eq.~\eqref{eq_gen_GF_dimless}, yields the supplementary conditions
\begin{align}
 f^\mu(k) & = g^\mu(-k) \,, \\[5pt]
 h(k) & = h(-k) \,,
\end{align}
which further restrict the gauge freedom of the Green function.

We remark that even in its full generality, Eq.~\eqref{mink_trans} is sufficient to guarantee the gauge invariance of the ``physical'' fields. To see this, we note that in Fourier space the four-potential is given in terms of ``its'' current density by Eq.~\eqref{eq_introD_0}, while on the other hand the continuity equation \eqref{eq_continuity} reads
\begin{equation}
 k_\mu \h j^\mu(k) = 0\,.
\end{equation}
Hence, applying the Green function on a ``physical'' current actually yields 
\begin{equation}
 A^\mu(k) = \mathbbmsl D_0(k) \left(\eta\indices{^\mu_\nu} + \frac{c k^\mu}{\omega} \h f_\nu(k)\right)j^\nu(k) \,. \label{eq_actually}
\end{equation}
Next, performing the transition to the field strength tensor by means of
\begin{equation}
 F^{\mu\nu} = \i \h (k^\mu A^\nu - k^\nu A^\mu) 
\end{equation}
shows that the second term in Eq.~\eqref{eq_actually} also cancels out and we arrive at the overall expression
\begin{equation}
 F^{\mu\nu}(k) = \i \h \mathbbmsl D_0(k) \, (k^\mu j^\nu(k) - k^\nu j^\mu(k)) \,,
\end{equation}
in which none of the arbitrary functions $f,g$ or $h$ appears. This shows that, as matter of principle,
one could have chosen from the very outset the Green function in the simple form
\begin{equation} \label{green_feyn}
 (D_0)\indices{^\mu_\nu}(k) = \mathbbmsl D_0(k) \, \eta\indices{^\mu_\nu} \,,
\end{equation}
without any effect on physically observable quantities. Therefore, 
the question arises of why one does not simply stick to this trivial choice of vanishing parameter functions?
The answer is of a rather technical nature: The Green function in the form of Eq.~\eqref{green_feyn} corresponds to the so-called ``Feynman gauge'' \cite[Eq.~(3.270)]{Bertlmann}, which in particular
implies the Lorenz gauge on the level of the four-potential, i.e., $k_\mu A^\mu=0$. Thus, by generally choosing  \mbox{$f=g=h=0$}, one \linebreak would be tied down to a special gauge condition,
while in actual fact there are definitely situations where other gauges are more practical.

Finally, it is instructive to compare the above expression \eqref{eq_gen_GF_dimless} of the Minkowski tensor $D_0$ with the decomposition  introduced in \S\,\ref{sec_proj}. In fact, Eq.~\eqref{eq_gen_GF_dimless} is equivalent to
\begin{equation}
 D_0(k) = D_{0, \hh \rm LL}(k) + D_{0, \hh \rm LT}(k) + D_{0, \hh \rm TL}(k) + D_{0, \hh \rm TT}(k) \,,
\end{equation}
where the four contributions are given by
\begin{align}
 D_{0, \hh \rm LL}(k) & = \mathbbmsl D_0(k) \h \bigg( 1 + \frac{c^2 k^2}{\omega^2} \h h(k) \bigg) P_{\rm L}(k) \,, \label{dnodll} \\[5pt]
 D_{0, \hh \rm TT}(k) & = \mathbbmsl D_0(k) \h P_{\rm T}(k) \,, \label{dnodtt}
\end{align}
as well as
\begin{align}
  (D_{0, \hh \rm LT})\indices{^\mu_\nu}(k) & = \mathbbmsl D_0(k) \, \frac{c k^\mu}{\omega} \h f_\nu(k) \,, \label{dnodlt} \\[3pt]
  (D_{0, \hh \rm TL})\indices{^\mu_\nu}(k) & = \mathbbmsl D_0(k) \, g^\mu(k) \h \frac{c k_\nu}{\omega} \,. \label{dnodtl}
\end{align}
For later purposes, we now rewrite these results in the $(3+1)$-formalism.

\subsubsection{(3+1)-formalism}

In the $(3+1)$-formalism of \S\,\ref{3plus1}, the constraints \eqref{mink_trans} imply that
\begin{equation} \label{vor_zwischen_6}
 f_\nu = \big({-\vec u} \cdot \vec f, \h \vec f^{\rm T}\hh\big) = \vec f^{\rm T} \h \big( {-\vec u}, \h \tsr 1 \hh \big) \,,
\end{equation}
and
\begin{equation} \label{zwischen_6}
 g^\mu = \bigg( \!\! \begin{array}{c} \vec u \cdot \vec g \\[3pt] \vec g \end{array} \!\! \bigg) = \bigg( \!\mh \begin{array}{c} \vec u^{\rm T} \\[2pt] \tsr 1 \end{array} \!\mh \bigg) \, \vec g \,. \smallskip
\end{equation}
Combining Eqs.~\eqref{zwischen_5} and \eqref{vor_zwischen_6}, we obtain
\begin{equation}
 \frac{c k^\mu}{\omega} \h f_\nu = \left( \!\! \begin{array}{cc} -\vec u \cdot \vec f & \vec f^{\rm T} \\[6pt] -(\vec u \cdot \vec f) \hh \vec u & \vec u \hh \vec f^{\rm T} \end{array} \!\! \right) \,.
\end{equation}
Similarly evaluating all other terms in Eq.~\eqref{eq_gen_GF_dimless}, we arrive at
\begin{equation} \label{eq_gen_GF_dimless_var}
 (D_0)\indices{^\mu_\nu} = \mathbbmsl D_0 \left( \! \begin{array}{cc} 1 - \vec u \cdot \vec f - \vec u \cdot \vec g - h & \vec f^{\rm T} + (\vec u \cdot \vec g) \h \vec u^{\rm T} + h \h \vec u^{\rm T} \\[6pt]
 -(\vec u \cdot \vec f) \h \vec u - \vec g - h \h \vec u & \tsr 1 + \vec u \hh \vec f^{\rm T} + \vec g \h \vec u^{\rm T} + h \h \vec u \vec u^{\rm T} \end{array} \! \right),
\end{equation}
with the scalar Green function given by Eq.~\eqref{defscalarD}, or equivalently by
\begin{equation}
 \mathbbmsl D_0 \equiv \mathbbmsl D_0(\vec u, \omega) = \frac{1}{\varepsilon_0 \h \omega^2} \h \frac{1}{|\vec u|^2 - 1} \,.
\end{equation}
We can also write this result as
\begin{align}
 (D_0)\indices{^\mu_\nu} & = \mathbbmsl D_0 \, \bigg\{ \bigg( \! \begin{array}{ll} 1 & 0 \\[3pt] 0 & \tsr 1 \end{array} \! \bigg) \\ \nonumber
 & \quad \, + \begin{pmatrix} 1 \\[2pt] \vec u \end{pmatrix} \mh \vec f^{\rm T} \big({-\vec u}, \h \tsr 1 \h \big) + \bigg( \!\mh \begin{array}{c} \vec u^{\rm T} \\[2pt] \tsr 1 \end{array} \!\mh \bigg) \h\hh \vec g \h\hh \big({-1}, \h \vec u^{\rm T} \hh \big) + h \begin{pmatrix} 1 \\[2pt] \vec u \end{pmatrix} \mh \begin{pmatrix} -1, \h \vec u^{\rm T} \hh \end{pmatrix} \!\bigg\} \,.
\end{align}
The seven component functions given by $\vec f = (f_1, \h f_2, \h f_3)^{\rm T}$, $\vec g = (g_1, \h g_2, \h g_3)^{\rm T}$ and $h$ can be chosen arbitrarily, 
and each choice yields a tensorial Green function which satisfies the defining Eq.~\eqref{eq_defEqFGF}. In particular, we may consider special cases where
\begin{align}
 \vec f(\vec u) & = f(\vec u) \h \vec u \,, \label{scalar_f} \\[3pt]
 \vec g(\vec u) & = g(\vec u) \h \vec u \,, \label{scalar_g}
\end{align}
with {\itshape scalar} functions $f$ and $g$. Then, Eq.~\eqref{eq_gen_GF_dimless_var} simplifies to
\begin{equation} \label{eq_gen_GF_dimless_simpl}
 (D_0)\indices{^\mu_\nu} = \mathbbmsl D_0 \left( \! \begin{array}{cc} 1 - (f + g) \h |\vec u|^2  - h & (f + g \h |\vec u|^2 + h) \h \vec u^{\rm T} \\[5pt]
 -(f \h |\vec u|^2 + g + h) \h \vec u & \tsr 1 + (f + g + h) \h \vec u \vec u^{\rm T} \end{array} \!\! \right).
\end{equation}
This form of the electromagnetic Green function is particularly suitable for the recovery of the temporal gauge, as we will show in the next subsection.

\subsection{Temporal gauge} \label{subsec_temp_gauge}

It has been shown already in Ref.~\cite[Eqs.~(3.57)--(3.59)]{ED1} that the free Green function in the temporal gauge can be obtained by choosing 
the scalar functions $f, \h g$ and $h$ as follows:
\begin{equation}
 f(\vec u) = \frac{1}{|\vec u|^2 - 1}\,, \qquad g(\vec u) = 0 \,, \qquad h(\vec u) = -f(\vec u) \,.
\end{equation}
In fact, by putting these functions into Eq.~\eqref{eq_gen_GF_dimless_simpl} we obtain
\begin{equation}
 (D_0)\indices{^\mu_\nu} = \mathbbmsl D_0 \left( \!\! \begin{array}{cc} 0 & 0 \\[5pt] -\vec u & \tsr 1 \end{array} \!\mh \right), \label{GF_Mink_temp}
\end{equation}
or in terms of the original variables $\vec k$ and $\omega$,
\begin{equation} \label{eq_weylgreen}
 (D_0)\indices{^\mu_\nu}(\vec k, \omega) = \mathbbmsl D_0(\vec k, \omega) \left( \!\! \begin{array}{cc} 0 & 0 \\[5pt]
 -c \hh \vec k / \omega & \tsr 1
 \end{array} \!\mh \right).
\end{equation}
As shown in Ref.~\cite[\S\,2.2.1]{EDWave}, this formula can also be derived directly from the equation of motion for the electromagnetic vector potential in the temporal gauge.
Either way, we refer to Eq.~\eqref{eq_weylgreen} as the free electromagnetic Green function in the {\itshape Minkowskian temporal gauge.}

Furthermore, we have derived in Ref.~\cite[\S\,2.2.2]{EDWave} an alternative form of the free electromagnetic Green function by replacing in the equation of motion the charge density with the current density via the continuity equation. Thus, the free electromagnetic Green function can be brought into the form
\begin{equation} \label{eq_weylgreen1}
 (D_0)\indices{^\mu_\nu}(\vec k, \omega) = \left(  \!
 \begin{array}{cc} 0 & 0 \\[5pt]
 0 & \tsr D_0(\vec k, \omega)
 \end{array} \!\mh \right),
\end{equation}
with the {\itshape free Cartesian Green function}
\begin{equation}\label{eq_GFtempGauge}
\tsr D_0(\vec k,\omega)=\mathbbmsl D_0(\vec k,\omega)\left(\tsr 1-\frac{c^2|\vec k|^2}{\omega^2}\tsr P\L(\vec k)\right),
\end{equation}
which can also be written compactly as
\begin{equation} \label{CartGreen_u}
 \tsr D_0 = \mathbbmsl D_0 \, (\tsr 1 - \vec u \vec u^{\rm T} \hh) \,.
\end{equation}
The above form \eqref{eq_weylgreen1} of the free electromagnetic Green function can in turn be obtained from the general expression \eqref{eq_gen_GF_dimless_simpl} by choosing
\begin{equation} \label{cart_temp_gauge}
 f(\vec u) = g(\vec u) = \frac{1}{|\vec u|^2 -1} \,, \qquad h(\vec u) = \frac{1 + |\vec u|^2}{1 - |\vec u|^2} \,.
\end{equation}
Correspondingly, we refer to Eq.~\eqref{eq_weylgreen1} as the free electromagnetic Green function in the {\itshape Cartesian temporal gauge.}

\section{Fundamental response tensor} \label{sec_fund_resp}
\subsection{Definition} \label{subsec_def_chi}

Within the limits of {\it linear response theory}, the basic quantity of electro\-{}dynamics in media is the {\it fundamental response tensor} defined as \cite[\S\,5.1]{ED1}
\begin{equation}
 \chi\indices{^\mu_\nu}(x,x') = \frac{\delta j^\mu\ind(x)}{\delta A^\nu\ext(x')} \,,
\end{equation}
such that the {\it induced} four-current can be expanded to linear order in terms of the {\it external} four-potential as
\begin{equation}
 j^\mu\ind(x) = \int \!\de^4 x'\, \chi\indices{^\mu_\nu}(x,x') \h A^\nu\ext(x') \,. \label{eq_expansion}
\end{equation}
The continuity equation for the induced four-current and its invariance under gauge transformations of the external four-potential imply the constraint relations
\begin{align}
 \partial_\mu \h \chi\indices{^\mu_\nu}(x,x') &= 0 \,,  \label{eq_constraint1} \\[5pt]
 \partial'_\nu \h \chi\indices{^\mu_\nu}(x,x') &= 0 \,. \label{eq_constraint2}
\end{align}
In the homogeneous limit, response functions depend only on the coordinate difference, $\chi(x,x')=\chi(x-x')$, 
such that the expansion \eqref{eq_expansion} can be written in Fourier space as a point-wise product,
\begin{equation}
 j^\mu\ind(k) = \chi\indices{^\mu_\nu}(k) \h A\ext^\nu(k) \,.
\end{equation}
In this limit, the constraints \eqref{eq_constraint1}--\eqref{eq_constraint2} read (cf.~Ref.~\cite[Eq.~(10.5.2)]{Weinberg1})
\begin{equation} \label{constr}
 k_\mu \h \chi\indices{^\mu_\nu}(k) = \chi\indices{^\mu_\nu}(k) \h k^\nu = 0 \,.
\end{equation}
These equations will become particularly important for the description of the fundamental response tensor in the $(3+1)$-formalism.

\subsection{General form} \label{fundresp_genform}

In terms of the Minkowskian longitudinal and transverse projectors of \S\,\ref{sec_proj},
the constraints \eqref{constr} can be written equivalently as
\begin{equation}
 P_{\rm L}(k) \h \chi(k) = \chi(k) P_{\rm L}(k) = 0 \,.
\end{equation}
This implies that in the general decomposition
\begin{equation} \label{gen_dec_chi}
 \chi = \chi_{\rm LL} + \chi_{\rm LT} + \chi_{\rm TL} + \chi_{\rm TT} \,,
\end{equation}
only the last term is nonzero, i.e.,
\begin{align}
 \chi_{\rm LL} = \chi_{\rm LT} = \chi_{\rm TL} & = 0  \,, \\[3pt]
 \chi_{\rm TT} & = \chi \,,
\end{align}
hence $\chi$ is a Minkowski-transverse tensor.
Moreover, it follows that in the $(3+1)$-formalism the fundamental response tensor attains the form
\begin{equation}
 \chi^\mu_{~\nu}=
\left( \!\! \begin{array}{rr} -\vec u^{\rm T} \h \tsr \chi \h \vec u & \vec u^{\rm T} \h \tsr \chi \\[5pt] -\tsr \chi \h \vec u & \tsr \chi \end{array} \! \right) .
\end{equation}
This can be written even more compactly as
\begin{equation} \label{chi_compact}
 \chi\indices{^\mu_\nu} = \bigg( \!\mh \begin{array}{c} \vec u^{\rm T} \\[2pt] \tsr 1 \end{array} \!\mh \bigg) \h \tsr \chi \, \big({-\vec u}, \h \tsr 1 \h \big) \,,
\end{equation}
where we have used the matrices defined in Eqs.~\eqref{def_mat_1}--\eqref{def_mat_2}.
In terms of the original variables $\vec k$ and $\omega$, we thus obtain the well-known representation of the fundamental response tensor (see Ref.~\cite{ED1} and references therein)
\begin{equation}\label{generalform1}
\chi^\mu_{~\nu}(\vec k,\omega)=
\left( \!\!
\begin{array}{rr} -\lar{\frac{c^2}{\omega^2}} \, \vec k^{\rm T} \, \tsr{\chi}(\vec k,\omega)\, \vec k & \lar{\frac{c}{\omega}} \, \vec k^{\rm T} \, \tsr{\chi}(\vec k,\omega)\, \\
[10pt] -\lar{{\frac{c}{\omega}}} \,\h \tsr{\chi}(\vec k,\omega)\, \vec k & \, \tsr{\chi}(\vec k,\omega)\,
\end{array} \! \right).
\end{equation}
In particular, this shows that the {\itshape current response tensor,} which is defined as the spatial part of the fundamental response tensor, already determines the whole fundamental response tensor and thus, ultimately, all linear electromagnetic response
properties (see Ref.~\cite[Sct.~6]{ED1}).


\subsection{Proper response tensor}

Similarly as the (``direct'') fundamental response tensor introduced in \S\,\ref{subsec_def_chi}, the {\it proper} fundamental response tensor is defined as the functional
derivative of the induced four-current with respect to the {\itshape total} (i.e., external plus induced) four-potential, i.e.,
\begin{equation}
 \widetilde\chi\indices{^\mu_\nu}(x,x') = \frac{\delta j^\mu\ind(x)}{\delta A^\nu\tot(x')} \,.
\end{equation}
Generally, proper response functions are relevant for the following reasons: (i) Within the Functional Approach to electrodynamics of media \cite{ED2, ED1}, they play a particularly important r\^{o}le in the derivation of wave equations in materials \cite{Refr,EDWave,OptTens,EDWegner}.
(ii) The Kubo formula as calculated by {\itshape ab initio} computer codes relying on independent-particle approximations should be interpreted as the respective proper response function \cite[\S\,II]{Schwalbe}.
Furthermore, as the analysis below will show, the proper response tensor is related to its ``direct'' counterpart by a Dyson equation. In a Feynman graph analy\-{}sis, this implies that the proper response tensor corresponds to a sum of irreducible graphs,  and hence it is more directly accessible by theoretical calculations.
(iii) In some cases, the commonly used response function has to be identified with the proper one anyway (for instance, ``the'' conductivity tensor in most cases actually refers to the proper conductivity tensor). Finally, we stress that the description of the response in terms of proper rather than ``direct'' response functions is at least possible and {\itshape a priori}
it is not clear why the latter approach should be favored over the former. Fittingly, 
the proper response tensor satisfies the same constraints as the fundamental response tensor, Eqs.~\eqref{eq_constraint1}--\eqref{eq_constraint2}, and is therefore also Minkowski-transverse,
\begin{equation}
 \widetilde \chi = \widetilde \chi\TT \,.
\end{equation}
Correspondingly, its general form in the homogeneous limit reads
\begin{equation}
 \widetilde \chi^\mu_{~\nu}=
\Bigg( \!\! \begin{array}{rr} -\vec u^{\rm T} \h \tsr{\widetilde \chi} \h \vec u & \vec u^{\rm T} \h \tsr{\widetilde \chi} \\[5pt] -\tsr{\widetilde \chi} \h \vec u & \tsr{\widetilde \chi} \end{array} \! \Bigg) \,.
\end{equation}
One can easily show by the functional chain rule that the fundamental response tensor is related to its proper counterpart by the Dyson-type equation
\begin{equation} \label{eq_Dyson}
 \chi = \widetilde \chi + \widetilde \chi \h D_0 \h \chi \,.
\end{equation}
We now draw a few direct conclusions from this equation. First, one shows easily that Eq.~\eqref{eq_Dyson} implies the relations
\begin{equation} \label{dys_impl_1}
 \chi = (1 - \widetilde \chi \h D_0)^{-1} \, \widetilde \chi \,,
\end{equation}
as well as \smallskip
\begin{equation} \label{dys_impl_2}
 1 + D_0 \h \chi = (1 - D_0 \h \widetilde \chi\hh)^{-1}  \,. \smallskip \vspace{2pt}
\end{equation}
Furthermore, since both $\chi$ and $\widetilde \chi$ are Minkowski-transverse tensors, we can replace $D_0$ in Eq.~\eqref{eq_Dyson} by its transverse projection, i.e.,
\begin{equation}
 \chi = \widetilde \chi + \widetilde \chi \, (D_0)\TT \, \chi \,,
\end{equation}
where $(D_0)\TT \equiv D_{0, \hh \rm TT}$ are two different notations for the same object. By means of the explicit expression \eqref{dnodtt}, this can be further simplified to
\begin{equation} \label{Dyson_reduced}
 \chi = \widetilde \chi + \mathbbmsl D_0 \h \widetilde \chi \h\hh \chi \,,
\end{equation}
with the scalar Green function $\mathbbmsl D_0$\h. From the ensuing formal expansion
\begin{equation}
 \chi \h = \h \widetilde \chi + \mathbbmsl D_0 \h \widetilde \chi^{\h2} + \mathbbmsl D_0^2 \, \widetilde \chi^{\h3} + \ldots  \h = \h \sum_{n = 1}^{\infty} \mathbbmsl D_0^{n-1} \h \widetilde \chi^{\h n} \,,
\end{equation}
we then deduce that the tensors $\chi$ and $\widetilde \chi$ commute with each other, i.e.,
\begin{equation}
 \chi \h\hh \widetilde \chi = \widetilde \chi \h\hh \chi \,.
\end{equation}
Moreover, the Dyson-type equation \eqref{Dyson_reduced} implies the following relations, which are analogous to Eqs.~\eqref{dys_impl_1}--\eqref{dys_impl_2}:
\begin{equation}
 \chi = X \h \widetilde \chi \,,
\end{equation}
with the dimensionless Minkowski tensor $X$ being given by
\begin{equation} \label{def_X}
 X := 1 + \mathbbmsl D_0 \h \chi = (1 - \mathbbmsl D_0 \h \widetilde \chi \hh )^{-1}  \,.
\end{equation}
Finally, by choosing the free electromagnetic Green function $D_0$ in the Cartesian temporal gauge (see \S\,\ref{subsec_temp_gauge}), the Dyson equation \eqref{eq_Dyson} translates into
\begin{align}
& \Bigg( \!\! \begin{array}{rr} -\vec u^{\rm T} \h \tsr \chi \h \vec u & \vec u^{\rm T} \h \tsr \chi \\[5pt] -\tsr \chi \h \vec u & \tsr \chi \end{array} \! \Bigg) = \Bigg( \!\! \begin{array}{rr} -\vec u^{\rm T} \h \tsr{\widetilde \chi} \h \vec u & \vec u^{\rm T} \h \tsr{\widetilde \chi} \\[5pt] -\tsr{\widetilde \chi} \h \vec u & \tsr{\widetilde \chi} \end{array} \! \Bigg) \\[3pt]
& + \Bigg( \!\! \begin{array}{rr} -\vec u^{\rm T} \h \tsr{\widetilde \chi} \h \vec u & \vec u^{\rm T} \h \tsr{\widetilde \chi} \\[5pt] -\tsr{\widetilde \chi} \h \vec u & \tsr{\widetilde \chi} \end{array} \! \Bigg)
 \left( \!
 \begin{array}{cc} 0 & 0 \\[5pt]
 0 & \tsr D_0
 \end{array}\! \right)
 \Bigg( \!\! \begin{array}{rr} -\vec u^{\rm T} \h \tsr \chi \h \vec u & \vec u^{\rm T} \h \tsr \chi \\[5pt] -\tsr \chi \h \vec u & \tsr \chi \end{array} \! \Bigg) \,.
\end{align}
Performing the matrix multiplications explicitly, we thus obtain
\begin{equation}
 \Bigg( \!\! \begin{array}{rr} -\vec u^{\rm T} \h \tsr \chi \h \vec u & \vec u^{\rm T} \h \tsr \chi \\[5pt] -\tsr \chi \h \vec u & \tsr \chi \end{array} \! \Bigg) =
 \left( \!\! \begin{array}{rr} -\vec u^{\rm T} \h \big(\tsr{\widetilde \chi} + \tsr{\widetilde \chi} \h\hh \tsr D_0 \h \tsr \chi \h \big) \h \vec u & \vec u^{\rm T} \h \big(\tsr{\widetilde \chi} + \tsr{\widetilde \chi} \h\hh \tsr D_0 \h \tsr \chi \h \big) \\[5pt] -\big(\tsr{\widetilde \chi} + \tsr{\widetilde \chi} \h\hh \tsr D_0 \h \tsr \chi\h\big) \h \vec u & \big(\tsr{\widetilde \chi} + \tsr{\widetilde \chi} \h\hh \tsr D_0 \h \tsr \chi\h\big) \end{array} \! \right),
\end{equation}
from which we read off the {\itshape Cartesian} Dyson-type equation
\begin{equation} \label{Cart_dyson_1}
 \tsr \chi = \tsr{\widetilde \chi} + \tsr{\widetilde \chi} \h\hh \tsr D_0 \h \tsr \chi \,,
\end{equation}
which is hence equivalent to its well-known Minkowskian version \eqref{eq_Dyson}.

\section{Full electromagnetic Green function} \label{sec_femGF}
\subsection{Definition}

The {\it free} electromagnetic Green function introduced in \S\,\ref{subsec_freeGF}
can be characterized as the functional derivative of the four-potential with respect to its own generating four-current,
\begin{equation}
 (D_0)\indices{^\mu_\nu}(x,x') = \frac{\delta A^\mu(x)}{\delta j^\nu(x')}\,.
\end{equation}
In the presence of materials, we introduce the analogous quantity \cite[\S\,5.2]{ED1}
\begin{equation}
 D\indices{^\mu_\nu}(x,x') = \frac{\delta A^\mu\tot(x)}{\delta j^\nu\ext(x')}	\,,
\end{equation}
which is correspondingly called the {\it full} electromagnetic Green function.
Using once more the functional chain rule, the full Green function can be related to its free counterpart and the fundamental response tensor as follows
(cf.~Ref.~\cite[Eq.~(10.5.10)]{Weinberg1}):
\begin{equation} \label{dyson_eq_exp}
D=D_0+D_0 \h \chi \h D_0\,. 
\end{equation}
Alternatively, the full Green function can be related to its free analogon and the {\it proper} fundamental response tensor by the {\itshape Dyson equation,} which reads as follows (cf.~Ref.~\cite[Eq.~(10.5.14)]{Weinberg1}):
\begin{equation} \label{dyson_eq}
D=D_0+D_0 \h \widetilde\chi \h D\,.
\end{equation}
Here, we have used the identities
\begin{equation}
 (D_0)\indices{^\mu_\nu}(x,x') = \frac{\delta A^\mu\ext(x)}{\delta j^\nu\ext(x')} = \frac{\delta A^\mu\ind(x)}{\delta j^\nu\ind(x')}\,.
\end{equation}
By iterating Eq.~\eqref{dyson_eq}, one obtains the perturbative series
\begin{equation}
 D=D_0+D_0 \h \widetilde\chi \h D_0 + \ldots \,,
\end{equation}
commonly used in quantum electrodynamics \cite[Eq.~(13.32)]{Huang}. 
Conceptually, the direct and the proper current response tensor respectively correspond to the reducible and the irreducible photon self-energy used in quantum electrodynamics (see Ref.~\cite[\S\,10.5]{Weinberg1}). 
Therefore, Eq.~\eqref{dyson_eq_exp} shows that
if one was in possession of the {\it full} current response tensor (i.e., the exact, reducible photon self-energy ``including all vertex corrections''), then one could also calculate the full electromagnetic Green function.
Unfortunately, in this case the {\itshape caveat} applies that the full current response tensor as calculated in condensed matter physics---even 
if directly and exactly so from the Kubo formula \cite[App.~C]{ED2} applied to the interacting many-electron system---would still not coincide
with its counterpart used in quantum electrodynamics \cite[\S\,10.5]{Weinberg1}. The reason for this is that 
in condensed matter physics---apart from the fact that one uses a nonrelativistic formulary---one restricts oneself to a purely electronic problem taking into
account only the Coulomb interaction, while in quantum electrodynamics one is obliged to treat the complete system of electrons and ``photons''.
Thus, a Feynman graph expansion of the current response tensor within quantum electrodynamics  would again involve the electromagnetic Green function,
which is not the case in condensed matter physics. Hence, from a practical point of view, 
if one is only given the ``full'' current response tensor in the sense of condensed matter physics,
it would still not be possible to calculate the full electromagnetic Green function from Eq.~\eqref{dyson_eq_exp}.
Apart from this {\itshape proviso,} however, the Dyson equation---although derived within the context of condensed matter physics---is indeed formally 
identical to the self-consistent equation for the full Green function used in quantum electrodynamics, meaning in particular that
the ``full'' electromagnetic Green function calculated in condensed matter physics can be regarded at a reasonable approximation
to its counterpart in quantum electro\-{}dynamics.

Be it as it may, with the above equations we can finally specify the principal problem treated in this article. In fact, as has been reviewed in Sct.\,\ref{sec_ClassED},
the solution of the equation of motion \eqref{eq_defEqFGF} for the free Green function is not uniquely determined.
Instead, its general form is given by Eq.~\eqref{eq_gen_GF_dimless}, which involves seven freely adjustable parameter functions
(namely, $h(k)$, $\vec f(k)$, and $\vec g(k)$). Consequently, the full Green function is not uniquely determined either.
The question therefore arises what {\itshape its} general form looks like, and how the adjustable parameters of the free electromagnetic
Green function translate into the corresponding adjustable parameters of the full Green function. These questions will be answered in the remaining part of this article.

\subsection{General form} \label{sec_genmink}
\subsubsection{Projector formalism}

The most general form of the full Green function in the homogeneous limit follows from the Dyson equation \eqref{dyson_eq}, which uniquely determines $D$ in terms of the free Green function $D_0$ and the proper fundamental response tensor $\widetilde \chi$ as a formal power series,
\begin{equation}
 D = D_0 + D_0 \h \widetilde \chi \h D_0 + D_0 \h \widetilde \chi \h D_0 \h \widetilde \chi \h D_0 + \ldots
\end{equation}
Even more straightforwardly, however, the general form of $D$ can be deduced from its representation \eqref{dyson_eq_exp} in terms of the free Green function~$D_0$ and the direct response tensor $\chi$\h. For this purpose, we use again the expansion of the full Green function,
\begin{equation}
 D = D_{\rm LL} + D_{\rm LT} + D_{\rm TL} + D_{\rm TT} \,,
\end{equation}
and  the analogous expansions of the free Green function and the fundamental response tensor (which have been  investigated in \S\,\ref{subsec_genformGF} and \S\,\ref{fundresp_genform}). Multiplying Eq.~\eqref{dyson_eq_exp} from left and right with $P_{\rm L}$ or $P_{\rm T}$, respectively, and using that $\chi$ is Minkowski-transverse, we obtain the four identities
\begin{align}
 D\LL & = (D_0)\LL + (D_0)\LT \,\h \chi \, (D_0)\TL \,, \label{dll} \\[5pt]
 D\LT & = (D_0)\LT + (D_0)\LT \,\h \chi \, (D_0)\TT \,, \label{dlt} \\[5pt]
 D\TL & = (D_0)\TL + (D_0)\TT \,\h \chi \, (D_0)\TL \,, \label{dtl} \\[5pt]
 D\TT & = (D_0)\TT + (D_0)\TT \,\h \chi \, (D_0)\TT \,. \label{dtt}
\end{align}
With the concrete expressions \eqref{dnodll}--\eqref{dnodtl} for the free Green function, we first obtain the longitudinal contribution of the full Green function,
\begin{align}
 & (D\LL)\indices{^\mu_\nu}(k) = \\[5pt] \nonumber
 & \mathbbmsl D_0(k) \, \bigg( \bigg( 1 + \frac{c^2 k^2}{\omega^2} \h h(k) \bigg) \h (P\L)\indices{^\mu_\nu}(k) + \mathbbmsl D_0(k) \, \frac{c k^\mu}{\omega} \, f_\lambda(k) \, \chi\indices{^\lambda_\rho}(k) \, g^\rho(k) \, \frac{c k_\nu}{\omega} \bigg) \,,
\end{align}
then the mixed contributions,
\begin{align}
 & (D\LT)\indices{^\mu_\nu}(k) = \mathbbmsl D_0(k) \h \bigg( \frac{c k^\mu}{\omega} f_\nu(k) + \mathbbmsl D_0(k) \, \frac{c k^\mu}{\omega} \h f_\lambda(k) \, \chi\indices{^\lambda_\nu}(k) \bigg) \,, \\[5pt]
 & (D\TL)\indices{^\mu_\nu}(k) = \mathbbmsl D_0(k) \h \bigg( g^\mu(k) \h \frac{c k_\nu}{\omega} + \mathbbmsl D_0(k) \, \chi\indices{^\mu_\rho}(k) \, g^\rho(k) \, \frac{c k_\nu}{\omega} \h \bigg) \,,
\end{align}
and finally the transverse contribution,
\begin{equation}
 (D\TT)\indices{^\mu_\nu}(k) = \mathbbmsl D_0(k) \h \Big( (P\T)\indices{^\mu_\nu}(k) + \mathbbmsl D_0(k) \h \chi\indices{^\mu_\nu}(k) \Big) \,.
\end{equation}
By summing up all contributions, we arrive at the following general form of the {\itshape full} electromagnetic Green function:
\begin{align} \label{eq_gen_full_GF_dimless}
 & 
D\indices{^\mu_\nu}(k) = \\[3pt] \nonumber
 & \mathbbmsl D_0(k) \mh \left( \mh X\indices{^\mu_\nu}(k) + \frac{c k^\mu}{\omega} \h F_\nu(k) + G^\mu(k) \h \frac{c k_\nu}{\omega} + \frac{c k^\mu}{\omega} \h H(k) \h \frac{c k_\nu}{\omega} \right),
\end{align}
where the dimensionless tensor $X$ was defined in Eq.~\eqref{def_X}, and where the dimensionless functions $F_\nu\hh, G^\mu, H$ are related to their counterparts $f_\nu\hh, \h g^\mu,h$ \linebreak in the analogous representation \eqref{eq_gen_GF_dimless} of the {\itshape free} electromagnetic Green function by the following equations:
\begin{align}
 F_\nu & = f_\lambda \h\hh X\indices{^\lambda_\nu} \,, \label{Ff} \\[6pt]
 G^\mu & = X\indices{^\mu_\rho}\h\hh g^\rho \,, \label{Gg} \\[4pt]
 H & = h + f_\lambda  \, ( X\indices{^\lambda_\rho}-\eta\indices{^\lambda_\rho} \hh ) \, g^\rho \,. \label{Hh}
\end{align}
One the one hand, the functions $F_\nu\hh, \h G^\mu, H$ can be regarded as parameter functions of the full Green function, which can be chosen arbitrarily up to the constraints of Minkowski-transversality (analogous to Eq.~\eqref{mink_trans}), i.e.,
\begin{equation}
 F_\nu(k) \h\hh k^\nu = k_\mu \h G^\mu(k) = 0 \,.
\end{equation}
On the other hand, these functions are uniquely determined once their counterparts $f_\nu\hh, \h g^\mu, h$ in the free Green function are fixed. Thus, we can also derive an explicit expression of  the full Green function in terms of the parameter functions of the free Green function: resubstituting the definition \eqref{def_X} of $X$ in Eqs.~\eqref{eq_gen_full_GF_dimless}--\eqref{Hh} and abbreviating $u^\mu = c k^\mu /\omega$, we obtain
\begin{equation} 
\begin{aligned}
 D\indices{^\mu_\nu} & = 
 \mathbbmsl D_0 \h \Big( \eta\indices{^\mu_\nu} + u^\mu f_\nu \, + g^\mu \h u_\nu + u^\mu \h h \, u_\nu \Big) \\[3pt]
 & \quad \, + \mathbbmsl D_0^2 \h\hh \Big( \chi\indices{^\mu_\nu} + u^\mu f_\lambda \, \chi\indices{^\lambda_\nu} + \chi\indices{^\mu_\rho}\, g^\rho \h u_\nu + u^\mu\h (f_\lambda \, \chi\indices{^\lambda_\rho} \, g^\rho\hh) \h u_\nu \Big) \,,
\end{aligned}
\end{equation}
or equivalently,
\begin{equation} \label{eq_gen_full_GF_dimless_fgh}
D\indices{^\mu_\nu} = (D_0)\indices{^\mu_\nu} + \mathbbmsl D_0^2 \, \big( \eta\indices{^\mu_\lambda} + u^\mu f_\lambda \hh\big) \h\hh \chi\indices{^\lambda_\rho} \h \big( \eta\indices{^\rho_\nu} + g^\rho \h u_\nu \big) \,,
\end{equation}
where the first term is just the free Green function given by Eq.~\eqref{eq_gen_GF_dimless}. In fact, this last equation can also be derived directly from Eq.~\eqref{dyson_eq_exp} by using the expression \eqref{eq_gen_GF_dimless} for the free Green function together with the constraints \eqref{constr} on the fundamental response tensor.

\subsubsection{(3+1)-formalism} \label{subsec_penultimate}

We now reformulate our results in the $(3+1)$-forma\-{}lism, which will turn out to be  useful for deriving the Cartesian Dyson equation in the following subsection.
First, the tensor $X$ defined by Eq.~\eqref{def_X} can be represented as
\begin{equation} \label{genform_X}
 X\indices{^\mu_\nu} =
\Bigg( \!\! \begin{array}{rr} 1 -\mathbbmsl D_0 \h\hh \vec u^{\rm T} \h \tsr{ \chi} \h \vec u & \mathbbmsl D_0 \h\hh \vec u^{\rm T} \h \tsr{ \chi} \\[5pt] -\mathbbmsl D_0 \h \tsr{ \chi} \h \vec u & \tsr 1 + \mathbbmsl D_0 \h \tsr{ \chi} \end{array} \! \Bigg) \,,
\end{equation}
which follows from the corresponding expression \eqref{generalform1} of the fundamental response tensor. In terms of the dimensionless matrix
\begin{equation} \label{def_X_spatial}
 \tsr X := \tsr 1 + \mathbbmsl D_0 \h \tsr \chi \,,
\end{equation}
which is the spatial part of the Minkowski tensor $X$, we obtain the equivalent representation
\begin{equation}
 X\indices{^\mu_\nu} = \Bigg( \!\! \begin{array}{cc} 1 - \vec u^{\rm T} (\tsr X - \tsr 1 ) \h \vec u & \vec u^{\rm T} (\tsr X - \tsr 1) \\[5pt] -(\tsr X - \tsr 1) \h \vec u & \tsr X \end{array} \!\! \Bigg) \,.
\end{equation}
Next, the Minkowski-transverse functions $F_\nu$ and $G^\mu$ defined in the previous subsection can be written as
\begin{equation}
 F_\nu = \big({-\vec u} \cdot \vec F, \h \vec F^{\rm T}\hh\big) \,,
\end{equation}
and \smallskip
\begin{equation}
 G^\mu = \bigg( \!\! \begin{array}{c} \vec u \cdot \vec G \\[3pt] \vec G \end{array} \!\mh \bigg) \,, \smallskip
\end{equation}
which is analogous to Eqs.~\eqref{vor_zwischen_6}--\eqref{zwischen_6}. By putting these expressions into Eq.~\eqref{eq_gen_full_GF_dimless} and performing the matrix multiplications, we arrive at the following expression of the full Green function in the $(3+1)$-formalism:
\begin{align}
 D\indices{^\mu_\nu} & = \mathbbmsl D_0 \, \Bigg( \!\! \begin{array}{cc} 1 - \vec u^{\rm T} (\tsr X - \tsr 1 ) \h \vec u & \vec u^{\rm T} (\tsr X - \tsr 1) \\[5pt] -(\tsr X - \tsr 1) \h \vec u & \tsr X \end{array} \!\! \Bigg) \\[5pt]
 & \quad \, + \mathbbmsl D_0 \left( \!\! \begin{array}{cc} - \vec u \cdot \vec F - \vec u \cdot \vec G - H & \vec F^{\rm T} + (\vec u \cdot \vec G) \h \vec u^{\rm T} + H \vec u^{\rm T} \\[6pt] \nonumber
 -(\vec u \cdot \vec F) \h \vec u - \vec G - H \vec u & \vec u \h \vec F^{\rm T} + \vec G \h \vec u^{\rm T} + H \vec u \vec u^{\rm T} \end{array} \! \right) .
\end{align}
Using once more the relation \eqref{def_X_spatial}, we can write this result equivalently as
\begin{align} \label{eq_gen_full_GF_dimless_var}
 D\indices{^\mu_\nu} & = \mathbbmsl D_0^2 \, \Bigg( \!\! \begin{array}{rr} -\vec u^{\rm T} \h \tsr{ \chi} \h \vec u & \vec u^{\rm T} \h \tsr{ \chi} \\[5pt] -\tsr{ \chi} \h \vec u & \tsr{ \chi} \end{array} \! \Bigg) \\[5pt]
 & \quad \, + \mathbbmsl D_0 \left( \! \begin{array}{cc} 1 - \vec u \cdot \vec F - \vec u \cdot \vec G - H & \vec F^{\rm T} + (\vec u \cdot \vec G) \h \vec u^{\rm T} + H \vec u^{\rm T} \\[6pt] \nonumber
 -(\vec u \cdot \vec F) \h \vec u - \vec G - H \vec u & \tsr 1 + \vec u \h \vec F^{\rm T} + \vec G \h \vec u^{\rm T} + H \vec u \vec u^{\rm T} \end{array} \! \right) ,
\end{align}
where the second term is now precisely analogous to Eq.~\eqref{eq_gen_GF_dimless_var} for the free Green function.

To proceed further, we express the parameter functions $\vec F$, $\vec G$ and $H$ in terms of their counterparts $\vec f$, $\vec g$ and $h$ appearing in the free Green function. Thus, we rewrite Eq.~\eqref{Ff} as
\begin{equation}
\begin{aligned}
 \big({-\vec u} \cdot \vec F, \h \vec F^{\rm T}\hh\big) & = \big({-\vec u} \cdot \vec f, \h \vec f^{\rm T}\hh\big) \\
  & \quad \, + \mathbbmsl D_0 \h \big({-\vec u} \cdot \vec f, \h \vec f^{\rm T}\hh\big) \, \Bigg( \!\! \begin{array}{rr} -\vec u^{\rm T} \h \tsr{ \chi} \h \vec u & \vec u^{\rm T} \h \tsr{ \chi} \\[5pt] -\tsr{ \chi} \h \vec u & \tsr{ \chi} \end{array} \! \Bigg) \,,
 \end{aligned}
\end{equation}
from which we obtain after a straightforward calculation,
\begin{equation} \label{res_Ff}
\vec F^{\rm T} = \vec f^{\rm T} + \mathbbmsl D_0 \, \vec f^{\rm T} \h (\tsr 1 - \vec u \vec u^{\rm T}) \, \tsr \chi \,.
\end{equation}
Similarly, Eq.~\eqref{Gg} implies that
\begin{equation} \label{res_Gg}
 \vec G = \vec g + \mathbbmsl D_0 \h\hh \tsr \chi \, (\tsr 1 - \vec u \vec u^{\rm T}) \, \vec g \,, \smallskip
\end{equation}
and from Eq.~\eqref{Hh} we obtain
\begin{equation} \label{res_Hh}
 H = h + \mathbbmsl D_0 \, \vec f^{\rm T} \h (\tsr 1 - \vec u \vec u^{\rm T}) \, \tsr \chi \, (\tsr 1 - \vec u \vec u^{\rm T}) \, \vec g \,.
\end{equation}
Putting these results into Eq.~\eqref{eq_gen_full_GF_dimless_var} yields after a lengthy but straightforward~calculation the following general expression of the full electromagnetic Green function:
\begin{equation} \label{full_result}
\begin{aligned}
 D\indices{^\mu_\nu} & = (D_0)\indices{^\mu_\nu} \\[2pt]
  & \quad \, + \mathbbmsl D_0^2 \, \Bigg( \!\! \begin{array}{c} \vec f_{\vec u}^{\rm T} \\[5pt] \tsr 1 - \vec u \vec u^{\rm T} + \vec u \vec f_{\vec u}^{\rm T} \end{array} \! \Bigg) \,\hh \tsr\chi \,\hh \big( \hh {-\vec g}_{\vec u}\hh, \tsr 1 - \vec u \vec u^{\rm T} + \vec g_{\vec u} \vec u^{\rm T} \h \big) \,,
\end{aligned}
\end{equation}
where we have introduced the functions
\begin{align}
 \vec f_{\vec u}(\vec u) & := \vec u + (\tsr 1 - \vec u \vec u^{\rm T}) \, \vec f(\vec u) \,, \label{def_fu} \\[1pt]
 \vec g_{\vec u}(\vec u) & := \vec u + (\tsr 1 - \vec u \vec u^{\rm T}) \, \vec g(\vec u)\,. \label{def_gu}
\end{align}
Note that the first term on the right-hand side of Eq.~\eqref{full_result} is just the free Green function, which as a $(4 \times 4)$-matrix is given in terms of $\vec f$, $\vec g$ and $h$ by Eq.~\eqref{eq_gen_GF_dimless_var}. By contrast, the second term depends on the current response tensor and formally involves the multiplication of a $(4\times 3)$-matrix, a $(3 \times 3)$-matrix and a $(3 \times 4)$-matrix.

We conclude this subsection with a few remarks concerning our result \eqref{full_result}. First, an even shorter derivation of this expression in the $(3+1)$-formalism can be given by starting from Eq.~\eqref{eq_gen_full_GF_dimless_fgh} of the previous subsection, which also relates the full Green function to the parameter functions of the free Green function (but in the manifestly Lorentz-covariant formalism). Expressing $f_\nu$, $g^\mu$ and $\chi\indices{^\mu_\nu}$ in terms of their respective spatial parts via Eqs.~\eqref{vor_zwischen_6}, \eqref{zwischen_6} and \eqref{chi_compact} leads directly to the result \eqref{full_result}.

Secondly, using the representation \eqref{CartGreen_u} of the free Cartesian Green function, we can rewrite Eqs.~\eqref{res_Ff}--\eqref{res_Hh} compactly as
\begin{align}
 \vec F^{\rm T} & = \vec f^{\rm T} \hh (\tsr 1 + \tsr D_0 \h \tsr \chi \hh ) \,, \\[5pt]
 \vec G & = (\tsr 1 + \tsr \chi \h \tsr D_0 ) \, \vec g\,, \\[5pt]
 H & = h + \mathbbmsl D_0^{-1} \h \vec f^{\rm T} \h \tsr D_0 \h \tsr \chi \h \tsr D_0 \, \vec g \,.
\end{align}
Correspondingly, the result \eqref{full_result} is equivalent to
\begin{equation}  \label{alt_result_D}
\begin{aligned}
 D\indices{^\mu_\nu} & = (D_0)\indices{^\mu_\nu} \\[2pt]
 & \quad \, + \Bigg( \! \begin{array}{c} \mathbbmsl D_0 \h \vec u^{\rm T} + \vec f^{\rm T} \hh \tsr D_0  \\[5pt] \mathbbmsl D_0  \tsr 1 + \vec u \h\vec f^{\rm T} \h \tsr D_0 \end{array} \! \Bigg) \,\hh \tsr\chi \,\hh \big( \hh {-\mathbbmsl D_0} \h \vec u - \tsr D_0 \, \vec g\h, \, \mathbbmsl D_0 \tsr 1 + \tsr D_0  \, \vec g \h \vec u^{\rm T} \h \big) \,.
\end{aligned}
\end{equation}
Finally, in the special case where $\vec f$ and $\vec g$ are parallel to $\vec u$ and given by Eqs.~\eqref{scalar_f}--\eqref{scalar_g}, the definitions \eqref{def_fu}--\eqref{def_gu} also simplify to
\begin{align}
 \vec f_{\vec u}(\vec u) & = f_{\vec u}(\vec u) \h \vec u \,, \\[5pt]
 \vec g_{\vec u} (\vec u) & = g_{\vec u}(\vec u) \h \vec u \,,
\end{align}
with the respective scalar functions being defined as
\begin{align}
 f_{\vec u}(\vec u) & = 1 + (1-|\vec u|^2) \h f(\vec u) \,, \label{eq_altf}	\\[5pt]
 g_{\vec u}(\vec u) & = 1 + (1-|\vec u|^2) \h g(\vec u) \,. \label{eq_altg}
\end{align}
This case will be important in the following subsection.

\subsection{Temporal gauge}\label{subsec_schluss}

It follows directly from Eq.~\eqref{cart_temp_gauge} that in the Cartesian temporal gauge,
\begin{equation}
 f_{\vec u}(\vec u) = g_{\vec u}(\vec u) = 0 \,.
\end{equation}
Thus, the general expression \eqref{full_result} of the full Green function simplifies to
\begin{align}
 D\indices{^\mu_\nu} & = \left( \!
 \begin{array}{cc} 0 & 0 \\[5pt]
 0 & \tsr D_0
 \end{array} \! \right) + \mathbbmsl D_0^2 \begin{pmatrix} 0 \\[5pt] \tsr 1 - \vec u \vec u^{\rm T} \end{pmatrix} \tsr \chi \, \big( 0, \h \tsr 1 - \vec u \vec u^{\rm T} \h \big) \\[5pt]
 & = \left( \!
 \begin{array}{cc} 0 & 0 \\[5pt]
 0 & \tsr D_0
 \end{array} \! \right) + \left( \!\! \begin{array}{c} 0 \\[5pt]
 \tsr D_0  \end{array} \!\mh \right) \tsr \chi \, \big( 0, \h \tsr D_0 \hh\big) \,,
\end{align}
where we have used again Eq.~\eqref{CartGreen_u}.
Multiplying out the matrices, we find that the full Green function is of the same form as the free Green function,
\begin{equation} \label{zwischen_12}
 D\indices{^\mu_\nu}  = \left( \!
 \begin{array}{cc} 0 & 0 \\[5pt]
 0 & \tsr D
 \end{array} \!\right) ,
\end{equation}
and the {\itshape full Cartesian Green function} is related to its free counterpart by
\begin{equation} \label{Cart_dyson_pre}
 \tsr D = \tsr D_0 + \tsr D_0 \h \tsr \chi \h \tsr D_0 \,.
\end{equation}
This is a Cartesian version of Eq.~\eqref{dyson_eq_exp}, which had been our starting point for deriving the general form of the full Green function. Together with the relation \eqref{Cart_dyson_1} between the direct and the proper current response tensor, Eq.~\eqref{Cart_dyson_pre} implies the corresponding {\itshape Cartesian Dyson equation}
\begin{equation} \label{Cart_dyson_2}
 \tsr D = \tsr D_0 + \tsr D_0 \h \tsr{\widetilde \chi} \, \tsr D \,,
\end{equation}
which constitutes a main result of this article. While we have derived Eqs. \eqref{Cart_dyson_pre} and \eqref{Cart_dyson_2} from the general form of the full Green function, these equations can also be deduced directly from their more fundamental Minkowskian \linebreak

\pagebreak \noindent
counterparts, Eqs.~\eqref{dyson_eq_exp} and \eqref{dyson_eq}, by using the free Green function in the Cartesian temporal gauge, Eq.~\eqref{eq_weylgreen1}, together with the general representation \eqref{generalform1} of the fundamental response tensor.

\section{Conclusion}

We have derived the most general form of the full electromagnetic Green function, both in the manifestly Lorentz-covariant formalism (see Eq.~\eqref{eq_gen_full_GF_dimless}), and in the $(3+1)$-formalism which is more suitable for applications in condensed matter physics (see Eq.~\eqref{eq_gen_full_GF_dimless_var}). These results generalize the corresponding expressions for the free electromagnetic Green function, Eqs.~\eqref{eq_gen_GF_dimless} and \eqref{eq_gen_GF_dimless_var}, which had been derived already in Ref.~\cite{ED1}. In particular, we have shown that the full Green function depends on seven complex parameter functions $(\vec F, \vec G, H)$, which can be chosen arbitrarily, but which are uniquely determined once the corresponding parameter functions $(\vec f, \vec g, h)$ of the free Green function are fixed. Thus, we could also express the full Green function in terms of the parameter functions of the free Green function, both in the manifestly Lorentz-covariant formalism (see Eq.~\eqref{eq_gen_full_GF_dimless_fgh}) and in the $(3+1)$-formalism (see Eqs.~\eqref{full_result}--\eqref{def_gu}, or Eq.~\eqref{alt_result_D}).

As a further more practical outcome of this analysis, we have derived Cartesian Dyson equations both for the full electromagnetic Green function (see Eqs.~\eqref{Cart_dyson_2}) and for the fundamental response tensor (see Eq.~\eqref{Cart_dyson_1}). These allow for a reduction of their original four-dimensional Lorentz formulation to a more economic three-dimensional formulation still being exact and hence encapsulating the complete information.

\section*{Acknowledgments}
This research was supported by the DFG grant HO 2422/12-1 and by the DFG RTG 1995. R.\,S. thanks the Institute for Theoretical Physics at TU Bergakademie Freiberg for its hospitality. We also thank the referee for helpful suggestions.

\begin{table}
\begin{center}
\renewcommand{\arraystretch}{2.0}
\begin{tabular}{p{4.0cm}p{4.0cm}}
\toprule[1pt]
\raisebox{2pt}{Minkowskian} & \raisebox{2pt}{Cartesian} \\
\midrule[1pt]
$\chi = \widetilde \chi + \widetilde \chi \h D_0 \h \chi$  & $\tsr \chi = \tsr{\widetilde \chi} + \tsr{\widetilde \chi} \h \tsr D_0 \h \tsr \chi$ \\[5pt]
\midrule
$D=D_0+D_0 \h \widetilde\chi \h D$  & $\tsr D = \tsr D_0 + \tsr D_0 \h \tsr{\widetilde \chi} \h \tsr D$ \\[5pt]
\midrule
$D=D_0+D_0 \h \chi \h D_0$  & $\tsr D = \tsr D_0 + \tsr D_0 \h \tsr{\chi} \h \tsr D_0$ \\[5pt]
\bottomrule[1pt]
\end{tabular}
\caption{Exact Dyson-type equations. \label{dyson_1}}

\vspace{2cm}
\captionsetup{width=9cm}
\begin{tabular}{p{4.0cm}p{4.0cm}}
\toprule[1pt]
\raisebox{2pt}{Cartesian} & \raisebox{2pt}{Scalar} \\
\midrule[1pt]
$\tsr \chi = \tsr{\widetilde \chi} + \tsr{\widetilde \chi} \h\hh \tsr D_0 \h \tsr \chi$ & $\upchi = \widetilde \upchi + \widetilde \upchi \h\hh v \h\hh \upchi$ \\[5pt]
\midrule
$(\tsr\varepsilon_{\rm r})^{-1} = \tsr 1 + \tsr D_0 \h \tsr{\chi}$ & $\varepsilon^{-1}_{\rm r}=1+v\h\hh \upchi$\\[5pt]
\midrule
$\tsr\varepsilon_{\rm r}=\tsr 1 - \tsr D_0 \h \tsr{\widetilde\chi}$ & $\varepsilon_{\rm r}= 1-v \h\hh \widetilde\upchi$ \\[5pt]
\bottomrule[1pt]
\end{tabular} \vspace{3pt}
\caption{Current response tensor and dielectric tensor (left column: exact relations, right column: relations between longitudinal response functions which are valid in the isotropic limit). \label{dyson_2}}
\end{center}
\end{table}

\begin{appendices}
\section{Cartesian, Minkowskian, and scalar equations}

In this appendix, we compare some  relations between Cartesian tensors, which were derived in this article, to their already well-known Minkowskian counterparts, as well as to their analogous {\itshape scalar} relations which are commonly used in electronic structure physics.

Concretely, Table \ref{dyson_1} summarizes the Dyson-type relations between the free electromagnetic Green function $D_0$\hh, the full electromagnetic Green function~$D$, the (direct) fundamental response tensor $\chi$\h, and the proper fundamental response tensor $\widetilde \chi$\h. Importantly, the more economic Cartesian equations are actually equivalent to their Minkowskian counterparts (provided that we choose the Cartesian temporal gauge, see \S\,\ref{subsec_temp_gauge}). The Cartesian equations correspond to Eqs.~\eqref{Cart_dyson_1}, \eqref{Cart_dyson_pre}, and \eqref{Cart_dyson_2} in the main text, whereas the Minkowskian versions are derived, for example, in Ref.~\cite[\S\,5.2]{ED1}.

Furthermore, Table \ref{dyson_2} summarizes relations between the dielectric tensor and the (direct or proper) current response tensor. In the homogeneous limit, where longitudinal and transverse response functions decouple, the general tensor relations can be reduced to simpler scalar relations, which are commonly employed in first-principles materials physics. These scalar relations involve the (direct) density response function $\upchi$\h, the proper density response function $\widetilde \upchi$\h, and the {\itshape longitudinal} dielectric function $\varepsilon_{\rm r}$\hh. In particular, in these formulae  the scalar  Coulomb kernel $v$ replaces the free Cartesian Green function. For a derivation of these relations, see \cite[\S\,3.4 and \S\,5.1]{EDWave}.

\end{appendices}

\bibliographystyle{model1-num-names}
\bibliography{masterbib}

\begin{thebibliography}{24}
\expandafter\ifx\csname natexlab\endcsname\relax\def\natexlab#1{#1}\fi
\providecommand{\url}[1]{\texttt{#1}}
\providecommand{\href}[2]{#2}
\providecommand{\path}[1]{#1}
\providecommand{\DOIprefix}{doi:}
\providecommand{\ArXivprefix}{arXiv:}
\providecommand{\URLprefix}{URL: }
\providecommand{\Pubmedprefix}{pmid:}
\providecommand{\doi}[1]{\href{http://dx.doi.org/#1}{\path{#1}}}
\providecommand{\Pubmed}[1]{\href{pmid:#1}{\path{#1}}}
\providecommand{\bibinfo}[2]{#2}
\ifx\xfnm\relax \def\xfnm[#1]{\unskip,\space#1}\fi
\bibitem[{Weinberg(1995)}]{Weinberg1}
\bibinfo{author}{S.~Weinberg}, \bibinfo{title}{{\itshape The quantum theory of
  fields}}, volume \bibinfo{volume}{1: {\itshape foundations}},
  \bibinfo{publisher}{Cambridge University Press},
  \bibinfo{address}{Cambridge}, \bibinfo{year}{1995}.
\bibitem[{Itzykson and Zuber(1980)}]{Itzykson}
\bibinfo{author}{C.~Itzykson}, \bibinfo{author}{J.-B. Zuber},
  \bibinfo{title}{{\itshape Quantum field theory}},
  \bibinfo{publisher}{McGraw-Hill, Inc.}, \bibinfo{address}{New York},
  \bibinfo{year}{1980}.
\bibitem[{Peskin and Schr\"oder(1995)}]{Peskin}
\bibinfo{author}{M.~E. Peskin}, \bibinfo{author}{D.~V. Schr\"oder},
  \bibinfo{title}{{\itshape An introduction to quantum field theory}},
  \bibinfo{publisher}{Addison-Wesley Publishing Company},
  \bibinfo{address}{Reading, MA}, \bibinfo{year}{1995}.
\bibitem[{Huang(2010)}]{Huang}
\bibinfo{author}{K.~Huang}, \bibinfo{title}{{\itshape Quantum field theory:
  from operators to path integrals}}, \bibinfo{edition}{2nd} ed.,
  \bibinfo{publisher}{{WILEY-VCH Verlag GmbH {\&} Co. KGaA}},
  \bibinfo{address}{Weinheim}, \bibinfo{year}{2010}.
\bibitem[{Ryder(1996)}]{Ryder}
\bibinfo{author}{L.~H. Ryder}, \bibinfo{title}{{\itshape Quantum field
  theory}}, \bibinfo{edition}{2nd} ed., \bibinfo{publisher}{Cambridge
  University Press}, \bibinfo{address}{Cambridge}, \bibinfo{year}{1996}.
\bibitem[{Hedin(1965)}]{Hedin65}
\bibinfo{author}{L.~Hedin},
\newblock \bibinfo{title}{{\itshape New method for calculating the one-particle
  {Green's} function with application to the electron-gas problem}},
\newblock \bibinfo{journal}{Phys. Rev.} \bibinfo{volume}{{\bfseries 139}}
  (\bibinfo{year}{1965}) \bibinfo{pages}{A796}.
\bibitem[{Hedin and Lundqvist(1969)}]{Hedin69}
\bibinfo{author}{L.~Hedin}, \bibinfo{author}{S.~Lundqvist},
\newblock \bibinfo{title}{{\itshape Effects of electron-electron and
  electron-phonon interactions on the one-electron states of solids}},
\newblock volume~\bibinfo{volume}{23} of
  \textit{\bibinfo{series}{\textnormal{Solid State Physics: Advances in
  Research and Applications}}}, \bibinfo{publisher}{Academic Press, Inc.},
  \bibinfo{address}{New York}, \bibinfo{year}{1969}, pp.
  \bibinfo{pages}{1--181}.
\bibitem[{Aryasetiawan and Gunnarsson(1998)}]{Aryasetiawan98}
\bibinfo{author}{F.~Aryasetiawan}, \bibinfo{author}{O.~Gunnarsson},
\newblock \bibinfo{title}{{\itshape The GW method}},
\newblock \bibinfo{journal}{Rep. Prog. Phys.} \bibinfo{volume}{{\bfseries 61}}
  (\bibinfo{year}{1998}) \bibinfo{pages}{237}.
\bibitem[{Friedrich and Schindlmayr(2006)}]{Schindl}
\bibinfo{author}{C.~Friedrich}, \bibinfo{author}{A.~Schindlmayr},
\newblock \bibinfo{title}{{\itshape Many-body perturbation theory: the GW
  approximation}},
\newblock in: \bibinfo{editor}{J.~Grotendorst}, \bibinfo{editor}{S.~Bl\"ugel},
  \bibinfo{editor}{D.~Marx} (Eds.), \bibinfo{booktitle}{{\itshape Computational
  nanoscience: do it yourself!}}, volume~\bibinfo{volume}{31} of
  \textit{\bibinfo{series}{\textnormal{NIC Series}}}, \bibinfo{publisher}{John
  von Neumann Institute for Computing}, \bibinfo{address}{J{\"u}lich},
  \bibinfo{year}{2006}, pp. \bibinfo{pages}{335--355}.
\bibitem[{Melrose(2008)}]{Melrose1Book}
\bibinfo{author}{D.~B. Melrose}, \bibinfo{title}{{\itshape Quantum
  plasmadynamics: unmagnetized plasmas}}, volume \bibinfo{volume}{735} of
  \textit{\bibinfo{series}{\textnormal{Lecture Notes in Physics}}},
  \bibinfo{publisher}{Springer}, \bibinfo{address}{New York},
  \bibinfo{year}{2008}.
\bibitem[{Starke and Schober(2016)}]{EDOhm}
\bibinfo{author}{R.~Starke}, \bibinfo{author}{G.~A.~H. Schober},
\newblock \bibinfo{title}{{\itshape Relativistic covariance of {Ohm's} law}},
\newblock \bibinfo{journal}{Int. J. Mod. Phys. D} \bibinfo{volume}{{\bfseries
  25}} (\bibinfo{year}{2016}) \bibinfo{pages}{1640010}. \bibinfo{note}{{See
  also arXiv:1409.3723 [math-ph]}}.
\bibitem[{Starke and Schober(2017)}]{EDLor}
\bibinfo{author}{R.~Starke}, \bibinfo{author}{G.~A.~H. Schober},
\newblock \bibinfo{title}{{\itshape Covariant response theory and the boost
  transform of the dielectric tensor}},
\newblock \bibinfo{journal}{Int. J. Mod. Phys. D} \bibinfo{volume}{{\bfseries
  26}} (\bibinfo{year}{2017}) \bibinfo{pages}{1750163}. \bibinfo{note}{{See
  also arXiv:1702.06985 [physics.class-ph]}}.
\bibitem[{Starke and Schober(2016)}]{ED2}
\bibinfo{author}{R.~Starke}, \bibinfo{author}{G.~A.~H. Schober},
  \bibinfo{title}{{\itshape Ab initio materials physics and microscopic
  electrodynamics of media}}, \bibinfo{howpublished}{arXiv:1606.00445
  [cond-mat.mtrl-sci]}, \bibinfo{year}{2016}.
\bibitem[{Starke and Schober(2017{\natexlab{a}})}]{Refr}
\bibinfo{author}{R.~Starke}, \bibinfo{author}{G.~A.~H. Schober},
\newblock \bibinfo{title}{{\itshape Microscopic theory of the refractive
  index}},
\newblock \bibinfo{journal}{Optik} \bibinfo{volume}{{\bfseries 140}}
  (\bibinfo{year}{2017}{\natexlab{a}}) \bibinfo{pages}{62}. \bibinfo{note}{{See
  also arXiv:1510.03404 [cond-mat.mtrl-sci]}}.
\bibitem[{Starke and Schober(2017{\natexlab{b}})}]{EDWave}
\bibinfo{author}{R.~Starke}, \bibinfo{author}{G.~A.~H. Schober},
\newblock \bibinfo{title}{{\itshape Linear electromagnetic wave equations in
  materials}},
\newblock \bibinfo{journal}{Phot. Nano. Fund. Appl.}
  \bibinfo{volume}{{\bfseries 26}} (\bibinfo{year}{2017}{\natexlab{b}})
  \bibinfo{pages}{41}. \bibinfo{note}{{See also arXiv:1704.06615
  [cond-mat.mtrl-sci]}}.
\bibitem[{Starke and Schober(2015)}]{ED1}
\bibinfo{author}{R.~Starke}, \bibinfo{author}{G.~A.~H. Schober},
\newblock \bibinfo{title}{{\itshape Functional Approach to electrodynamics of
  media}},
\newblock \bibinfo{journal}{Phot. Nano. Fund. Appl.}
  \bibinfo{volume}{{\bfseries 14}} (\bibinfo{year}{2015})
  \bibinfo{pages}{1--34}. \bibinfo{note}{{See also arXiv:1401.6800
  [cond-mat.mtrl-sci]}}.
\bibitem[{Misner et~al.(1973)Misner, Thorne, and Wheeler}]{Misner}
\bibinfo{author}{C.~W. Misner}, \bibinfo{author}{K.~S. Thorne},
  \bibinfo{author}{J.~A. Wheeler}, \bibinfo{title}{{\itshape Gravitation}},
  \bibinfo{publisher}{W. H. Freeman and Company}, \bibinfo{address}{San
  Francisco}, \bibinfo{year}{1973}.
\bibitem[{Alcubierre(2008)}]{Alcubierre}
\bibinfo{author}{M.~Alcubierre}, \bibinfo{title}{{\itshape Introduction to
  $3+1$ numerical relativity}}, volume \bibinfo{volume}{140} of
  \textit{\bibinfo{series}{\textnormal{International Series of Monographs in
  Physics}}}, \bibinfo{publisher}{Oxford University Press},
  \bibinfo{address}{Oxford}, \bibinfo{year}{2008}.
\bibitem[{Gambini and Pullin(2011)}]{GambiniPullin}
\bibinfo{author}{R.~Gambini}, \bibinfo{author}{J.~Pullin},
  \bibinfo{title}{{\itshape A first course in Loop Quantum Gravity}},
  \bibinfo{publisher}{Oxford University Press}, \bibinfo{address}{Oxford},
  \bibinfo{year}{2011}.
\bibitem[{Starke and Schober(2016)}]{EffWW}
\bibinfo{author}{R.~Starke}, \bibinfo{author}{G.~A.~H. Schober},
  \bibinfo{title}{{\itshape Response Theory of the electron-phonon coupling}},
  \bibinfo{howpublished}{arXiv:1606.00012 [cond-mat.mtrl-sci]},
  \bibinfo{year}{2016}.
\bibitem[{Bertlmann(1996)}]{Bertlmann}
\bibinfo{author}{R.~A. Bertlmann}, \bibinfo{title}{{\itshape Anomalies in
  quantum field theory}}, \bibinfo{publisher}{Oxford University Press Inc.},
  \bibinfo{address}{New York}, \bibinfo{year}{1996}.
\bibitem[{Starke et~al.(2017)Starke, Schober, Wirnata, and Kortus}]{OptTens}
\bibinfo{author}{R.~Starke}, \bibinfo{author}{G.~A.~H. Schober},
  \bibinfo{author}{R.~Wirnata}, \bibinfo{author}{J.~Kortus},
  \bibinfo{title}{{\itshape Wavevector-dependent optical properties from
  wavevector-independent conductivity tensor}},
  \bibinfo{howpublished}{arXiv:1708.06330 [physics.optics]},
  \bibinfo{year}{2017}.
\bibitem[{Schober and Starke(2018)}]{EDWegner}
\bibinfo{author}{G.~A.~H. Schober}, \bibinfo{author}{R.~Starke},
\newblock \bibinfo{title}{{\itshape Microscopic theory of refractive index
  applied to metamaterials: effective current response tensor corresponding to
  standard relation $n^2 = \varepsilon_{\textnormal{eff}} \hspace{1pt}
  \mu_{\textnormal{eff}}$}},
\newblock \bibinfo{journal}{Eur. Phys. J. B} \bibinfo{volume}{{\bfseries 91}}
  (\bibinfo{year}{2018}) \bibinfo{pages}{146}. \bibinfo{note}{{See also
  arXiv:1709.08811 [physics.class-ph]}}.
\bibitem[{Schwalbe et~al.(2016)Schwalbe, Wirnata, Starke, Schober, and
  Kortus}]{Schwalbe}
\bibinfo{author}{S.~Schwalbe}, \bibinfo{author}{R.~Wirnata},
  \bibinfo{author}{R.~Starke}, \bibinfo{author}{G.~A.~H. Schober},
  \bibinfo{author}{J.~Kortus},
\newblock \bibinfo{title}{{\itshape Ab initio electronic structure and optical
  conductivity of bismuth tellurohalides}},
\newblock \bibinfo{journal}{Phys. Rev. B} \bibinfo{volume}{{\bfseries 94}}
  (\bibinfo{year}{2016}) \bibinfo{pages}{205130}.

\end{thebibliography}

\end{document}